\documentclass[copyright,creativecommons]{eptcs}
\providecommand{\event}{SCSS 2021} 
\usepackage{underscore}           
\usepackage{bm}
\usepackage{booktabs}   
\usepackage{subcaption} 

\usepackage[utf8]{inputenc} 
\usepackage[T1]{fontenc}    
\usepackage{hyperref}
\usepackage{url}            
\usepackage{booktabs}       
\usepackage{amsfonts}       
\usepackage{nicefrac}       

\usepackage{graphicx} 
\usepackage{pgfplots} 
\usepackage{float}
\usepackage{bussproofs}
\usepackage{xspace}
\usepackage{amsmath}
\usepackage{amssymb}
\usepackage{mathtools}
\usepackage{alltt}

\usepackage{tikz} 
\usetikzlibrary{arrows,shapes}
\usetikzlibrary{trees,positioning,fit}
\tikzstyle{block} =
  [rectangle, draw,
   minimum width=9em, text centered, rounded corners, minimum height=3.5em]
\tikzstyle{line}=[draw]
\tikzstyle{cloud} =
  [draw, text centered, ellipse, minimum height=3.5em, minimum width=9em]

\usetikzlibrary{arrows,shapes,calc}
\usetikzlibrary{trees,positioning,fit}
\tikzstyle{arrow}=[draw,-to,thick]
\tikzstyle{bluearrow}=[draw,-to,thick,blue]
\tikzstyle{tcircle} = [circle, draw, fill=white]
\tikzstyle{tsquare} = [rectangle, draw, fill=white]
\tikzstyle{bcircle} = [circle, draw, blue, fill=blue]
\tikzstyle{gcircle} = [circle, draw, mygreen, fill=mygreen]

\def\holfour{\textsf{HOL4}\xspace}
\def\isabellehol{\textsf{Isabelle/HOL}\xspace}
\def\hollight{\textsf{HOL Light}\xspace}

\def\coq{\textsf{Coq}\xspace}

\def\eprover{\textsf{E prover}\xspace}

\def\tactictoe{\textsf{TacticToe}\xspace}
\usepackage{color}

\usetikzlibrary{arrows,shapes,calc}
\usetikzlibrary{trees,positioning,fit}
\tikzstyle{arrow}=[draw,-to,thick]
\tikzstyle{embedding} = [draw, minimum width=15mm, minimum height=6mm]
\tikzstyle{nnop} = [draw, minimum width=8mm, minimum height=8mm, rounded 
corners]
\tikzstyle{nnopbig} = [draw, minimum width=15mm, minimum height=15mm, rounded 
corners]
\tikzstyle{block} =
  [rectangle, draw,
   minimum width=9em, text centered, rounded corners, minimum height=2.0em]
\tikzstyle{line}=[draw]
\tikzstyle{cloud} =
  [draw, text centered, ellipse, minimum height=2.0em, minimum width=9em]

\usepackage[]{listings}
\lstdefinelanguage{SML}
{columns=fixed,
  keywords={THEN,THENL,val,open,store_thm,fun,fn,let,in,end,true,
  while,do,if,then,else,break,return,;},%
  frame=none,
  sensitive=true,
  keywordstyle=\fontfamily{lmss}\scriptsize\selectfont,%
  basicstyle=\fontfamily{pcr}\small\selectfont,%
  stringstyle=\tt,
  morestring=[b]",
  literate=
   {>>}{{\tt{>\hspace{0mm}>}}}1%
   {>|}{{\tt{>|}}}1%
   {=}{{\tt\raisebox{-.15mm}{=}}}1%
   {[}{{\tt\raisebox{-.15mm}{[}}}1%
   {]}{{\tt\raisebox{-.15mm}{]}}}1%
   {->}{{$\rightarrow$}}1%
   {wedge}{{$\wedge$}}1%
   {vee}{{$\vee$}}1%
   {==>}{{$\Rightarrow$}}1%
   {<==>}{{$\Leftrightarrow$}}1%
   {vee}{{$\vee$}}1%
   {<=}{{$\leq$}}1%
   {>=}{{$\geq$}}1%
   {'a}{{$\alpha$}}1%
   {'b}{{$\beta$}}1%
   {ldots}{$\ldots$}1%
   {emptyset}{$\emptyset$}1%
   {!}{$\forall$}1%
   {fff}{{DB.fetch}}1
  }

\makeatletter
\renewcommand\section{\@startsection{section}{1}{\z@}%
                       {-12\p@ \@plus -4\p@ \@minus -4\p@}%
                       {8\p@ \@plus 4\p@ \@minus 4\p@}%
                       {\normalfont\large\bfseries\boldmath
                        \rightskip=\z@ \@plus 8em\pretolerance=10000 }}
\makeatother

\begin{document}

\title{Learned Provability Likelihood\\ for Tactical Search\thanks{ 
Supported by the Czech Science Foundation project 20-06390Y}}
\author{Thibault Gauthier
\institute{Czech Technical University in Prague, Prague, Czech Republic}
\email{email@thibaultgauthier.fr}}
\def\authorrunning{T. Gauthier}
\def\titlerunning{Learned Provability Likelihood for Tactical Search}

\maketitle    
          
\begin{abstract}
We present a method to estimate the provability 
of a mathematical formula. We adapt the tactical theorem prover \tactictoe to 
factor in these estimations. Experiments over the \holfour library show an 
increase in the number of theorems re-proven by
\tactictoe thanks to this additional guidance. This amelioration in 
performance together with concurrent updates to the \tactictoe 
framework lead to an improved user experience.
\end{abstract}

\section{Introduction}
We take inspiration from the instinct shown by mathematicians
when attempting to prove a theorem. An important choice to make in their 
situation is whether to continue exploring a certain line of work or to 
switch to another approach entirely. 
The quality of this decision is influenced greatly by one's experience with 
other proof attempts. 
Our aim is to integrate such feedback to improve automation in interactive 
theorem provers (ITPs).
In the majority of ITPs, most proofs are build using tactics in a goal-oriented 
manner. A tactic is a procedure that takes as input a goal $g$ (a sequent in 
\holfour) and returns a list of goals whose conjunction implies $g$.
We can classify \holfour tactics into four categories.
Solvers attempts to prove the goal using general or domain-specific knowledge 
(e.g. $\mathit{metis\_tac}$~\cite{hurd05} for first-order logic). Rewrite 
tactics modify 
subterms of a goal by applying 
rewrite rules constructed from proven equalities,
Induction tactics (e.g. $\mathit{Induct}$) split the goal into 
multiple cases (typically a base case and an inductive case).
Kernel-level tactics (e.g. $\mathit{exist\_tac}$) allow for more refined 
control of the proof state.
Given these tactics, it is possible to create an automated prover that searches 
for the proof of a starting goal by predicting suitable tactics for 
each intermediate goal. In the following, we refer to such automation as a 
tactic-based automated theorem prover (ATP). Multiple tactic-based ATPs have 
been developed in the course of the last five years and are now one of the most 
effective~\cite{DBLP:journals/jar/GauthierKUKN21,DBLP:conf/mkm/BlaauwbroekUG20} 
general proof automation available in an ITP.
However, none of the tactic-based ATP so far take into
account the provability of the goals produced by each tactic.
Therefore, in this project, a tree neural network (TNN) is taught a 
function estimating the provabilty of goals in \holfour~\cite{hol4}. This 
\textit{value} function produces feedback signals called rewards that further 
guide the search algorithm of the tactic-based ATP
\tactictoe~\cite{DBLP:journals/jar/GauthierKUKN21}.

\paragraph{Related Works}
The most successful related tactic-based ATPs are
\textsf{Tactician}\xspace~\cite{DBLP:conf/mkm/BlaauwbroekUG20} for 
\coq~\cite{coq},
\textsf{Pamper}\xspace~\cite{DBLP:conf/kbse/NagashimaH18} for 
\isabellehol~\cite{isabelle},
and \textsf{HOList}\xspace~\cite{DBLP:conf/icml/BansalLRSW19} for 
\hollight~\cite{hollight}.
The \textsf{Tactician}\xspace is very user-friendly. In particular, 
it is the only one that can record tactic calls on the fly. 
It uses the $k$-nearest neighbor algorithm for tactic selection as \tactictoe 
does. However, it does not predict argument theorems independently of tactics.
In \textsf{Pamper}\xspace , the policy predictors are decision trees 
trained on top of human-engineered features.
In \textsf{HOList}\xspace , the prediction effort is 
concentrated on learning the policy for a few selected tactics and their 
arguments (theorems) using deep reinforcement learning.
A related field of research is machine learning for
first-order ATPs.
The \textsf{ENIGMA}\xspace~\cite{DBLP:conf/cade/ChvalovskyJ0U19} 
system for \eprover gathers positive and negative clauses from 
successful proof attempts. We use a similar technique to collect
our training examples.
The ATP \textsf{LeanCoP}\xspace~\cite{DBLP:conf/nips/KaliszykUMO18} 
has been trained via a reinforcement learning loop using 
boosted random forest predictors.
Its proof search relies on the same variant of Monte Carlo Tree 
Search (MCTS)~\cite{silver2017mastering} as \tactictoe. 
Nevertheless, it ignores the goal selection issue and thus could benefit 
directly from the MCTS adaptions proposed in this paper.

\section{Monte Carlo Tree Search with Tactics}
\begin{figure}
\resizebox{1.0\linewidth}{!}{

\begin{tikzpicture}[node distance=0.7cmn]
\node [] (0) {$\vdash \forall n.\ 2 \times \mathit{SUM}\ (n+1)\ I = n 
\times (n+1)$};
\node [below of=0,node distance=0.5cm] (0v) {22 $\rightarrow$  
23,\hspace{3mm} 
0.045 $\rightarrow$ 0.043};
\node [draw,fit=(0) (0v)] (0f) {};

\node [below of=0,node distance=2.5cm] (1) {$srw\_tac\ []\ X$};
\node [below of=1,node distance=0.5cm] (1v) {3,\hspace{3mm} 0,\hspace{3mm}  
0.25};
\node [draw,fit=(1) (1v)] (1f) {};
\draw[-to] (0f.south) to (1f);

\node [left of=1,node distance=6cm] (1l) {$Induct$};
\node [below of=1l,node distance=0.5cm] (1lv) {12 $\rightarrow$ 13,\hspace{3mm}
$8.7 \times 10^ {-7} \rightarrow 3.0 \times 10^ {-6}$,\hspace{3mm} 0.5};
\node [draw,fit=(1l) (1lv)] (1lf) {};
\draw[-to,ultra thick] (0f.south) to node[xshift=-50](t1){tactic selection} 
(1lf);

\node [right of=1,node distance=5cm] (1r) {$asm\_simp\_tac\ arith\_ss\ X$};
\node [below of=1r,node distance=0.5cm] (1rv) {2,\hspace{3mm} 0,\hspace{3mm} 
0.125};
\node [draw,fit=(1r) (1rv)] (1rf) {};
\draw[-to] (0f.south) to (1rf);

\node [below of=1l,node distance=3.5cm] (2a) 
  {$\vdash 2 \times \mathit{SUM}\ (0+1)\ I = 0 \times (0+1)$};
\node [below of=2a,node distance=0.5cm] (2ad) 
                                  {6 $\rightarrow$ 7,\hspace{3mm} $0.049 
                                  \rightarrow 0.18$};

\node [draw,fit=(2a) (2ad)] (2af) {};
\draw[-to,ultra thick] (1lf) to node[yshift=-10, xshift=38]{goal selection}  
(2af);

\node [right of=2a,node distance=9cm] (2b) 
  {$ 2 \times \mathit{SUM}\ (n+1)\ I = n \times (n+1)$};
\node [below of=2b,node distance=0.5cm] (2bd)
  {$\vdash  2 \times \mathit{SUM}\ (\mathit{SUC}\ n+1)\ I = \mathit{SUC}\ n 
\times (\mathit{SUC}\ n+1)$};
\node [below of=2bd,node distance=0.5cm] (2bdd) 
  {6,\hspace{3mm} $3.0 \times 10^ {-5}$};
\node [draw,fit=(2b) (2bd)(2bdd)] (2bf) {};
\node [draw,fit=(2af) (2bf)] (2f) {};
\draw[-to] (1lf.south) to (2bf);

\node [below of=2a,node distance=3.5cm] (3aa) {$\mathit{rewrite\_tac\ X}$};
\node [below of=3aa,node distance=0.5cm] (3aav) {$3 \rightarrow 4$,\hspace{3mm} 
$0.00021 \rightarrow 0.23$,\hspace{3mm} 0.5};
\node [draw,fit=(3aa) (3aav)] (3aaf) {};
\draw[-to,ultra thick] (2af.south) to node[yshift=-10,xshift=40](t3){tactic 
selection} (3aaf);

\node [right of=3aa,node distance=9cm] (3a) {$\mathit{srw\_tac\ [arith\_ss]\ 
 [SUM\_def, \ldots]}$};
\node [below of=3a,node distance=0.5cm] (3av) {2,\hspace{3mm} 0.15,\hspace{3mm} 
0.25};
\node [draw,fit=(3a) (3av)] (3af) {};
\draw[-to] (2af.south) to (3af);

\node [below of=3aa,node distance=3cm] (4aa) {$[\mathit{numeral\_distrib}]$};
\node [below of=4aa,node distance=0.5cm] (4aav) 
  {3,\hspace{3mm} $2.1 \times 10^{-4}$,\hspace{3mm} 0.5};
\node [draw,fit=(4aa) (4aav)] (4aaf) {};
\draw[-to] (3aaf.south) to (4aaf);

\node [right of=4aa,node distance=4.5cm] (4ab) {$[\mathit{MULT\_CLAUSES}]$};
\node [below of=4ab,node distance=0.5cm] (4abv) {1 $\rightarrow$ 
2,\hspace{3mm} 0 $\rightarrow$ 0.48,\hspace{3mm} 0.25};
\node [draw,dotted,fit=(4ab) (4abv)] (4abf) {};
\draw[-to, ultra thick] (3aaf.south) to  node[xshift=54]{argument selection} 
(4abf);

\node [below of=3a,node distance=2cm] (4a) {$\mathit{SUM}\ 1\ I = 0$};
\node [below of=4a,node distance=0.5cm] (4av) {1, 0.29};
\node [draw,fit=(4a) (4av)] (4af) {};
\node [draw,fit=(4af)] (4aff) {};
\draw[-to] (3af.south) to (4af);

\node [below of=4aa,node distance=2cm] (5a) {$2 \times 
\mathit{SUM}\ 1\ I = 0$};
\node [below of=5a,node distance=0.5cm] (5av) {2,\hspace{3mm} $3.1 \times 10 ^ 
{-4}$};
\node [draw,fit=(5a) (5av)] (5af) {};
\node [draw,fit=(5af)] (5aff) {};
\draw[-to] (4aaf.south) to (5af);

\node [below of=4ab,node distance=2.5cm] (5b) {$2 \times 
\mathit{SUM}\ (0 + 1)\ I = 0$};
\node [below of=5b,node distance=0.5cm] (5bv) 
{1,\hspace{3mm} 0.95};
\node [draw,dotted,fit=(5b) (5bv)] (5bf) {};
\node [draw,dotted,fit=(5bf)] (5bff) {};
\draw[-to,dotted] (4abf.south) to node[xshift=36]{node extension} (5bff);

\end{tikzpicture}
}
\caption{\label{fig:running} 
Iteration 22 of the value-guided MCTS loop on the goal
$\vdash \forall n.\ 2 \times \mathit{SUM}\ (n+1)\ I = n \times (n + 1)$.
The term $\mathit{SUM}\ n\  f$ stands for $\sum_{x=0}^{x < n} 
f (x)$ and $I$ is the identity function.
Each node contains in the following order: 
the number of visits with a possible update, the average of the 
rewards and the prior policy for tactics (and arguments). 
The presence of an arrow after these numbers indicates a backup update.
The selection path is made bold and created nodes are dotted. 
Saturated tactic (and argument) nodes and the subtree of 
the inductive case are omitted. Tactics may contain the placeholder $X$ 
~\cite{DBLP:journals/jar/GauthierKUKN21}
to indicate that an argument has to be provided.
To avoid dividing by 0 in the PUCT formula, 
all tactic (and argument) nodes are initialized 
with one visit and a reward of 0.
A proof is found after 427 iterations.
}
\end{figure}
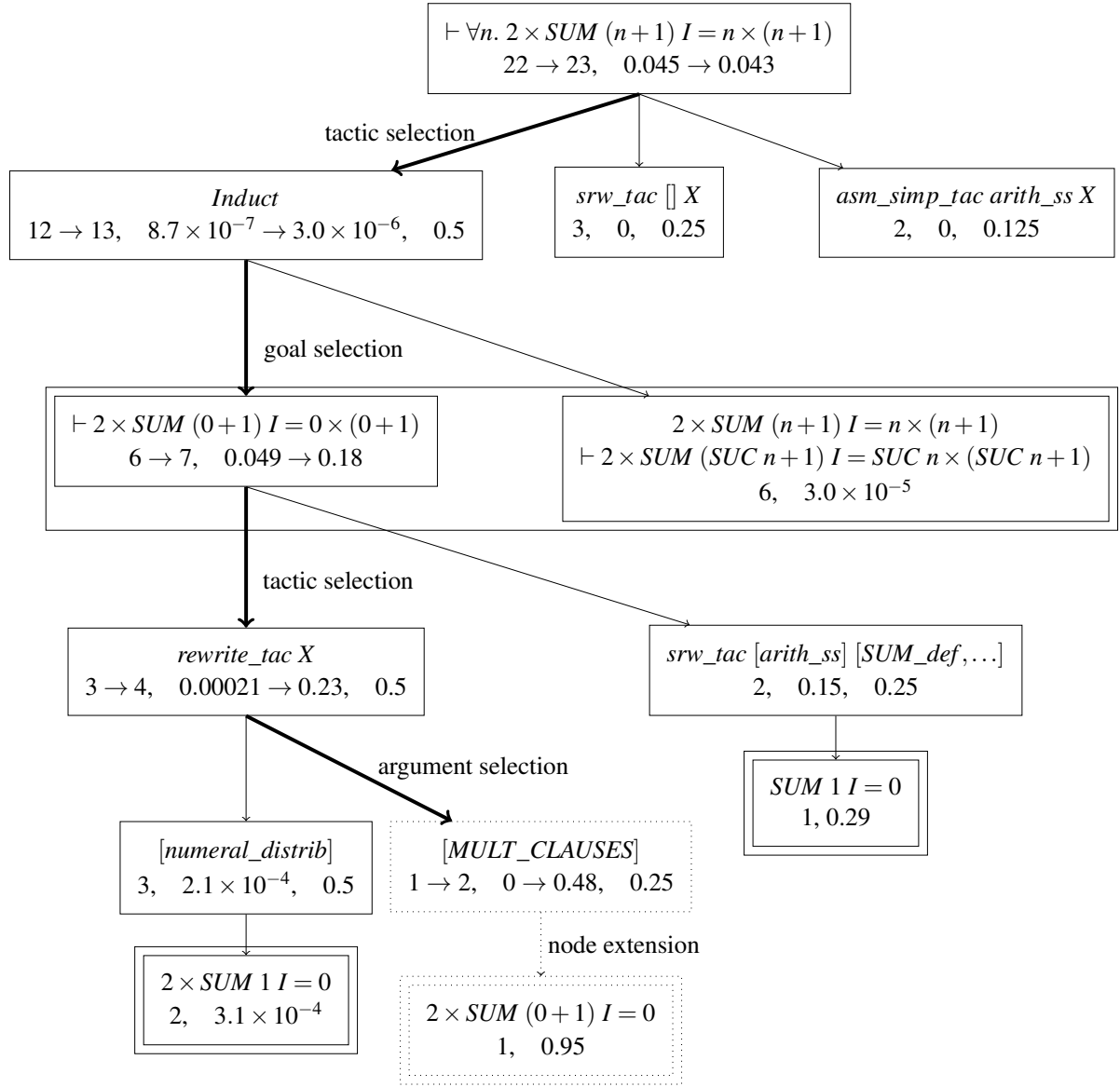

We integrate the learned provability estimator into the proof search of 
\tactictoe. Here is a brief summary of how the MCTS algorithm operates in the 
context of tactic-based theorem proving.
The algorithm starts with a list of goals (typically a singleton) to be 
proven in a root node.
In the selection phase, it chooses a goal 
branch in the current output node and a tactic branch (and possibly an argument 
branch) leading to the selection of an output node containing the list of goals 
produced by the tactic. This process is repeated until a leaf is reached.
In the extension phase, the tactic $t$ selected in the leaf is applied to its 
parent goal. In the case of a successful tactic application, an output leaf is 
created containing the list of goals produced by $t$. During the 
backup phase, a feedback 
signal is propagated from the newly created leaf to the root. The gathered 
node rewards influence the next selection phase. The three phases are repeated 
until the algorithm finds a proof for each of the root goals, times out, or 
saturates.
In the following, we present the existing MCTS algorithm for \tactictoe
and describe ways to improve the quality of the feedback 
mechanism.  Figure~\ref{fig:running} illustrates the effect of one iteration of 
the improved MCTS loop on the search tree for our running example.

%
%
%
%

\paragraph{Selection Phase} 
Given a list of tactic $(t_r)_{1 \leq r \leq n}$ with parent node $g$,
we can compute their PUCT~\cite{DBLP:conf/pkdd/AugerCT13} (Polynomial Upper 
Confidence Trees) score as follows:

\[PUCT (t_r) = \mathit{AverageRewards}(t_r) + 
  c \times \mathit{Policy}(t_r) \times 
  \frac{\sqrt{\mathit{Visits}(t_r)}}{\mathit{Visits}(g)}\]

The tactic with the highest PUCT score is selected.
The policy $\mathit{Policy}$ is given by a nearest neighbor predictor 
trained from supervised data consisting of goal-tactic 
pairs~\cite{DBLP:journals/jar/GauthierKUKN21}.
The value of $\mathit{Policy}(t_j)$ is experimentally chosen to be $0.5 ^ 
{n+1}$ where $n$ is the number of open tactic branches with parent 
node $g$ and higher nearest neighbor score. At the start, the search is 
principally guided towards nodes with higher policy. As the number of 
iterations increases, the search tends to explore
goals with higher reward averages more often. The exploration coefficient $c$ 
decides how fast this transition happens. We choose $c$ to be $2.0$ since this 
is a suitable value for guiding a goal-oriented 
first-order ATP~\cite{DBLP:conf/nips/KaliszykUMO18}.

In this version of \tactictoe, the theorems predicted by its nearest neighbor 
algorithm are split. Given a list of predicted arguments 
$x_1,\ldots,x_n$ and a tactic $t$ (except $\mathit{metis\_tac}$) that expects a 
list of theorems as argument, 
the tactic calls $t[x_1],\ldots,t[x_n]$ are constructed instead of 
$t[x_1,\ldots,x_n]$. 
This adds another branching layer to the tree that functions exactly like the 
tactic selection layer.
To simplify our explanations in the rest of this paper, 
the branching occurring during argument selection is considered to be part of 
tactic selection.

In previous developments of \tactictoe and in other tactic-based ATPs, the 
selected goal is always the first unproven goal of the output node. One issue 
is that the rewards of an output node are exactly the rewards of its first goal 
until it is proven. To factor the influence of other goals in the output node 
rewards, each unproven goal branch now receives almost the same number of 
visits. This is achieved by choosing at each iteration one of the least visited 
unproven goal branches.

\paragraph{Backup Phase}
There are three possible outcomes of the extension phase. 
If the tactic $t$ applied proves the parent goal $g$, $t$ receives a 
reward of 1. If $t$ fails or induces a loop then $t$ receives a reward of 0.
Otherwise, a new output node is created and the value network is called on each 
of the goals. The reward of the 
newly created leaf is computed by multiplying the inferred values and backed up 
unchanged to $t$. 
Each tactic selection layer propagates the reward of the selected tactic 
unchanged to its parent goal.
To capture the influence of multiple explored goal branches in the goal 
selection layer, we back up the rewards through this layer in the following 
manner.
Given a list of goals $(g_r)_{1 \leq r \leq n}$ composing an output node $p$ 
and a selected goal $g_i$, the reward for $p$ is given by:
   \[\mathit{Reward}_k (p) =  \mathit{Reward}_k (g_i) \times \prod_{
   1 \leq r \leq n\ \wedge\ r \neq i} 
   \mathit{AverageRewards}(g_r)\]

If a goal $g_r$ is proven, the value of 
$\mathit{AverageRewards}(g_r)$ in the formula is overridden and set to 1. 
This formula multiplies the reward of the goal $g_i$ at iteration $k$ of 
the search loop with the existing reward average $\mathit{AverageRewards}$ for 
the other goals in $p$. 
Thus the feedback of different branches is merged 
although only one branch is explored at each iteration of the loop.
Overall, $p$ receives a reward that is a lower bound
estimation for its provability as it assumes independence of its goals.

\section{Learning Provability}~\label{sec:method}
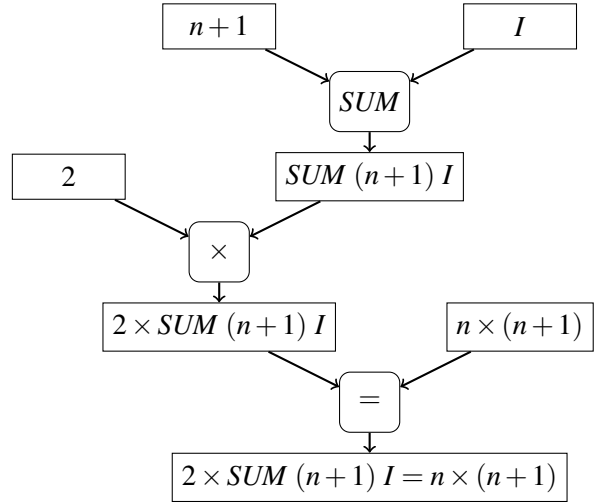
\begin{figure}
	\centering
	\vspace{-5mm}
	\begin{tikzpicture}[node distance=4cm]
    
    \node [embedding] (n+1) {$n + 1$};
    \node [right of=n+1,node distance=2cm] (n+1I) {};    
    \node [embedding,right of=n+1] (I) {$I$}; 
    \node [right of=I,node distance=2cm] (In) {};       
    \node [nnop,below of=n+1I,node distance=1cm] (sum) {$\mathit{SUM}$};    
    \node [embedding,below of=sum,node distance=1cm] (sumr) {$\mathit{SUM}\ 
    (n+1)\ I$};   
    \node [embedding,left of=sumr] (2) {$2$};
    \node [left of=sumr, node distance=2cm] (2m) {};
    \node [nnop,below of=2m,node distance=1cm] (times2) {$\times$};      
    \node [embedding,below of=times2,node distance=1cm] (times2r) {$ 2 
    \times 
    \mathit{SUM}\ (n+1)\ I$};      
    \node [right of=times2r,node distance=2cm] (times2rm) {};        
    \node [embedding, right of=times2r] (times2rr) {$n
    	\times (n + 1)$};  
    \node [nnop,below of=times2rm,node distance=1cm] (eq) {$=$};  
    \node [embedding,below of=eq,node distance=1cm] (eqr) {$2 \times 
    \mathit{SUM}\ (n+1)\ I = n \times (n + 1)$};  
    \draw[-to,thick] (n+1) to (sum);
    \draw[-to,thick] (I) to (sum);
    \draw[-to,thick] (sum) to (sumr);
    \draw[-to,thick] (2) to (times2);    
     \draw[-to,thick] (sumr) to (times2);  
     \draw[-to,thick] (times2) to (times2r);  
	 \draw[-to,thick] (times2r) to (eq);  
	 \draw[-to,thick] (times2rr) to (eq);  
	    \draw[-to,thick] (eq) to (eqr);  
	 \end{tikzpicture}
	\caption{Computation of the embedding of $
        2 \times \mathit{SUM}\ (n+1)\ I = n 
		\times (n + 1)$ by a TNN. 
		To simplify the diagram, the 
		computation of the 
		embeddings of $1$, $2$, $n+1$, $n \times (n+1)$ are not 
		shown. Rectangles represent embeddings and squares represent neural 
		networks.}\label{fig:tnn}.
\end{figure}

We explain here how a TNN can be trained to estimate the provability of a goal.
These estimates are to be used as rewards in our improved MCTS algorithm.
We choose a TNN as our machine learning model because it performs well 
on arithmetic and propositional formulas~\cite{DBLP:conf/mkm/Gauthier20} as well
as on Diophantine equations and combinators~\cite{DBLP:conf/lpar/Gauthier20}.
In our TNN, each \holfour operator of arity $a$ has a neural network
associated with it modeling a function from $\mathbb{R}^{a \times d}$ to 
$\mathbb{R}^{d}$, where $d$ is a globally fixed embedding size. When $a = 0$, 
the 
associated neural network is a trainable embedding 
(vector in $\mathbb{R}^{d}$). Networks are composed in a manner that reflects 
the tree structure of a given goal $g$. They gradually construct embeddings for 
sub-trees of $g$.
A head network models a function from $\mathbb{R}^{d}$ to $\mathbb{R}$ 
targeting a provability estimation for $g$ from 
an embedding for $g$.
Figure~\ref{fig:tnn} shows part of the computation graph
produced by a TNN on the running example.

Until now, the value function of \tactictoe was uniform.
It gave a reward of 1 independently of the provability of the goal produced. 
Thus, the search was guided solely by a policy trained from human-written proof 
scripts. The lack of negative examples in these proofs makes this 
dataset unsuitable for training the value function.
That is why we extract training examples from \tactictoe searches instead.
To do so, we collect search trees from the successful proof attempts.
The goal nodes extracted from these trees create our training examples 
with proven goals labeled positively and other goals negatively.
In order to speed up the training process, we keep at most one example for each 
goal with a preference for the positive ones. 
We also remove examples with large goals ($\geq 80$ 
operators) and select the $600$ most visited negative examples per proof 
attempt. From this dataset, we train a TNN with embedding size ($d=16$) and one 
layer 
per operator. The training is scheduled to take 100 epochs at a 
learning rate of 0.08 and the batch size is increased regularly according to 
the sequence $[16,24,32,48,64]$. These parameters were optimized by  
experimenting with a fifth of the \holfour library. There, the parameters used 
in ~\cite{DBLP:conf/mkm/Gauthier20} were gradually mutated to yield higher 
accuracy on \holfour goals.

\section{Results}~\label{sec:exp}
We experiment on the \holfour library included in the \holfour 
repository~\footnote{\url{https://github.com/HOL-Theorem-Prover/HOL}}.
This library contains 168 theories about mathematical and computer science 
concepts such as lists, trees, probabilities, measures and integration.
During the development of \holfour library,
the proof of a theorem is usually found as an argument of one of the following 
functions:
\texttt{store\_thm},\texttt{maybe\_thm},
\texttt{Store\_thm},\texttt{asm\_store\_thm},\texttt{prove} or 
\texttt{TAC\_PROOF}.
We introduce a hook in these functions in order to evaluate \tactictoe on
their argument goal as the library is built. 
In this way, only tactics, theorems and simplification sets created before a 
goal $g$ is proven are 
available to \tactictoe when attempting to prove $g$.
The code for this experiment is available at this 
commit\footnote{13fbff7b94bb1a5b51881a5aeee63ffc44d3be1}.  
Replication instructions are given in the file
src/tactictoe/EVALUATION.

After an evaluation of \tactictoe with a 30 seconds timeout over 15948 theorems
of the library resulting in 8812 successful searches,
188409 examples (34021 positives and 154388 negatives) are collected.
Correcting for the imbalance between the positive and negative examples by 
oversampling positive examples did not seem to improve the overall performance 
when experimenting with a small part of the library. Therefore, no balancing 
method is applied in this full-scale experiment.
The efficiency of the learning phase is then assessed by splitting the 
examples into a training set (90\%) and a test set (10\%).
Following the training of the TNN, we measure an accuracy of 
97.5\% on 
the training set and 84.7\% on the test set.
For the final evaluation, the TNN is retrained on all examples. 

The performance of
value-guided \tactictoe is compared with its default version in 
Table~\ref{tab:results}. All changes to the proof search algorithm proposed in 
this paper are included in both versions. The only difference between the two 
versions is whether a uniform or trained value function gives the reward signal.
Overall, the results show a small increase in the number of theorems re-proven.
Calls to the TNN increase the node creation 
time from 3.7\% to 13.3\% of the total search time. Such a small footprint 
could not have been achieved without a Standard ML implementation of TNNs 
~\cite{DBLP:conf/mkm/Gauthier20}. Thus, 
the positive impact of learning is not severely dampened by the 
neural network overhead.

\begin{table}
\centering
\setlength{\tabcolsep}{3mm}
\begin{tabular}{@{}lccc@{}}
\toprule
 &  \tactictoe & $\tactictoe_{\mathit{TNN}}$ & Combined \\
\midrule 
Standard library (15948 theorems) & 8812 (138) & 8913 (239) & 
9051 \\
\bottomrule
\end{tabular}
\vspace{2mm}
\caption{\label{tab:results} Number of theorems in the \holfour standard 
library re-proven within 30 seconds.
The number of theorems re-proven by one strategy but not by the other is
shown in parentheses.
}
\vspace{-5mm}
\end{table}

\subsection{Proofs}
Among the 239 theorems solely proven by $\tactictoe_{\mathit{TNN}}$, 30 of 
them belong to the theory real\_topology.
Two such theorems with their tactical proof are analyzed here.
Compared to the previous version of \tactictoe, the proofs are now 
automatically printed with the preferred style of many \holfour users. 
Tactics are written in lowercase when possible and the tacticals 
\texttt{>\hspace{0mm}>} (one goal), \texttt{>|} (multiple goals)
compose tactics in the final proof script.

The theorem $\vdash \forall a.\ \mathit{diameter}\ 
\{a\} = 0$ is re-proven in 
28.2 seconds. It states that the diameter of any singleton is equal to 0. The 
diameter of a set of reals $s$ is by definition $0$ if $s$ is empty and 
otherwise is the 
supremum $\{\mathit{abs} (x - y)\ |\ x \in s \wedge y \in s\}$.
The proof starts by rewriting the goal with the definition of 
$\mathit{diameter}$, continues by distinguishing whether $s$ is empty or 
not and proves each case by calling a first-order solver and a simplification 
tactic with an appropriate lemma.


\begin{lstlisting}[language=SML]
  rewrite_tac [diameter] >> REPEAT strip_tac >> 
  COND_CASES_TAC >| [metis_tac [], srw_tac [] [REAL_SUP_UNIQUE]]
\end{lstlisting}

The theorem $\vdash \forall f\ s.\ 
(\mathit{linear}\ f \wedge 
\mathit{subspace}\ s) \Rightarrow \mathit{subspace}\ (\mathit{IMAGE}\ f\ s)$ 
is re-proven in 16.7 seconds. It states that
the image of a subspace by a linear function is a subspace. A subspace $s$ is a
subset of $\mathbb{R}$ that satisfies three conditions $0 \in s$, $(x \in s 
\wedge y \in s) \Rightarrow x 
+ y \in s$, and 
$\forall c\in \mathbb{R}.\ x \in s \Rightarrow c \times x \in 
s$ 
.
The proof consists of rewriting the goal with this definition and then solving
each of the three cases by calling $\mathit{metis\_tac}$ with suitable premises.

\begin{lstlisting}[language=SML]
  srw_tac [] [subspace] >| 
    [metis_tac [REAL_MUL_LZERO,linear], 
     first_assum (X_CHOOSE_TAC ``B`` o MATCH_MP (LINEAR_BOUNDED)) >> 
       pop_assum (mp_tac o Q.SPEC `x`) >> metis_tac [LINEAR_ADD], 
     metis_tac [LINEAR_CMUL]]
\end{lstlisting}

\section{Usage}
To make \tactictoe more attractive to new 
users, tactic-goal pairs from the \holfour library are prerecorded. 
Therefore, the users can now use \tactictoe right after building \holfour 
without spending multiple hours recording the data themselves (see 
HOL/src/tactictoe/README).
Multiple new functions relying on the TNN trained in this paper 
are implemented and available at the commit mentioned in 
Section~\ref{sec:exp}\ . To run these functions, one first need to download
the file \texttt{tnn\_for\_tactictoe} from 
\url{http://grid01.ciirc.cvut.cz/~thibault/} to 
a desired location \texttt{path\_to\_foo}. 
The function \texttt{ttt\_tnn} runs \tactictoe$_\mathit{TNN}$ on a chosen goal
with the advice of a TNN imported by \texttt{mlTreeNeuralNetwork.read_tnn}.
The following commands demonstrate how to execute \texttt{ttt\_tnn} in an 
interactive session:

\begin{alltt}
load "tacticToe"; open tacticToe;
val tnn = mlTreeNeuralNetwork.read_tnn "path_to_foo";
load "sum_numTheory"; open sum_numTheory; open arithmeticTheory; 
set_timeout 60.0;
ttt_tnn tnn ([],\`{}\`{}!n. 2 * SUM (n+1) I = n * (n+1) \`{}\`{});
\end{alltt}

The function $\texttt{confidence\_tnn}$ returns the provability estimate of a 
goal as computed by the TNN in 1 to 10 milliseconds. This might encourage the 
user to call \tactictoe$_\mathit{TNN}$ if this value becomes high on an open 
goal during a manual proof attempt.

The function $\texttt{suggest}$ creates partial proofs from failed proof 
attempts. 
By limiting the numbers of iterations of the MCTS loop to 22 
(\texttt{tttSearch.looplimit := SOME 22;}), a call to \texttt{ttt\_tnn} now 
fails on the running example but the search tree is stored in the reference 
$\texttt{searchtree\_glob}$.
Executing $\texttt{suggest}\ ();$ at this point returns the most promising 
partial proof tree in $\texttt{searchtree\_glob}$.
To construct the tree, this function traverses down the tree and selects the 
first proved tactic branch or the most visited open tactic branch if no branch 
has been proved.
A tactic built from this proof tree can then be 
applied to the original goal and the user might continue working on the 
produced open goals. 
The resulting partial proof suggested from our failed attempt is:

\begin{lstlisting}[language=SML]
Induct >| [rewrite_tac [numeralTheory.numeral_distrib] >> all_tac,  
          srw_tac [ARITH_ss] [MULT_CLAUSES,LEFT_ADD_DISTRIB] >> all_tac]
\end{lstlisting}

The depth of this advice may be controlled by updating the reference 
$\texttt{tttSearch.suggest\_depth}$. The function $\texttt{suggest}$ can also 
be used in combination with the original \texttt{ttt} function but the advice 
will be of lesser value as the most promising proof branches are fixed by the 
policy and therefore the most promising proof tree is gradually extended as the 
search tree grows. 
In contrast, the decision on which path is 
most promising might be revised when running \texttt{ttt\_tnn} with 
higher timeouts.

\section{Conclusion}
Training a TNN on intermediate goals generated by 
\tactictoe leads to the creation of a goal provability estimator 
with good accuracy. Since a goal can be split into multiple goals during 
tactic-based searches, we propose improvements to the selection and backup 
phases of the underlying MCTS 
algorithm in order to get more accurate feedback from the learned estimator.
Experiments on the full \holfour library measure an increase in the 
success rate of \tactictoe and demonstrate the scalability of our approach.
This resulted in the creation of new tools for HOL4 users that leverage
the advice given by the estimator.
In particular, \tactictoe is now able to suggest how to start a proof 
even when it fails to find one.

\bibliographystyle{eptcs}
\bibliography{biblio}

\end{document}